\documentclass[reprint,amsmath,amssymb,aps,prl,superscriptaddress]{revtex4-1}

\usepackage{microtype}
\AtBeginEnvironment{verbatim}{\microtypesetup{activate=false}}

\usepackage{graphicx}
\usepackage{dcolumn}
\usepackage{bm}
\usepackage[pdftex,colorlinks=true,linkcolor=blue,citecolor=blue]{hyperref}
\usepackage[T1]{fontenc}
\usepackage{physics}
\usepackage{braket}
\usepackage[table, usenames, dvipsnames]{xcolor}
\usepackage{makecell}
\usepackage{boldline}
\setcellgapes{3pt}
\usepackage{placeins}

\usepackage{textgreek}

    \usepackage{lineno}  
    \usepackage{amsmath} 
    \usepackage{etoolbox} 
    \newcommand*\linenomathpatch[1]{%
    \cspreto{#1}{\linenomath}%
    \cspreto{#1*}{\linenomath}%
    \csappto{end#1}{\endlinenomath}%
    \csappto{end#1*}{\endlinenomath}%
    }
    \linenomathpatch{equation}
    \linenomathpatch{gather}
    \linenomathpatch{multline}
    \linenomathpatch{align}
    \linenomathpatch{alignat}
    \linenomathpatch{flalign}

\usepackage[strict]{changepage}
\usepackage{nicefrac}

\begin{document}

\title{Giant non-reciprocity and gyration through modulation-induced Hatano-Nelson coupling in integrated photonics}

\author{Oğulcan E. Örsel}
\affiliation{Department of Electrical $\&$ Computer Engineering, University of Illinois at Urbana–Champaign, Urbana, IL 61801 USA}

\author{Jiho Noh}
\affiliation{Sandia National Laboratories, Albuquerque, NM 87185, USA.}
\affiliation{Department of Mechanical Science $\&$ Engineering, University of Illinois at Urbana–Champaign, Urbana, IL 61801 USA}

\author{Penghao Zhu}
\affiliation{Department of Physics, The Ohio State University, Columbus, OH 43210, USA}
\affiliation{Department of Physics,  University of Illinois at Urbana–Champaign, Urbana, IL 61801 USA}

\author{Jieun Yim}
\affiliation{Department of Mechanical Science $\&$ Engineering, University of Illinois at Urbana–Champaign, Urbana, IL 61801 USA}
 
\author{Taylor L. Hughes}
\affiliation{Department of Physics,  University of Illinois at Urbana–Champaign, Urbana, IL 61801 USA}
 
\author{Ronny Thomale}
\affiliation{Institute for Theoretical Physics and Astrophysics, University of Würzburg, D-97074 Würzburg, Germany}
 
\author{Gaurav Bahl}
\affiliation{Department of Mechanical Science $\&$ Engineering, University of Illinois at Urbana–Champaign, Urbana, IL 61801 USA}

\date{\today}

\begin{abstract}
Asymmetric energy exchange interactions, also known as Hatano-Nelson type couplings, enable the study of non-Hermitian physics and associated phenomena like the non-Hermitian skin effect and exceptional points (EP). 
Since these interactions are by definition non-reciprocal, there have been very few options for real-space implementations in integrated photonics. In this work, we show that real-space asymmetric couplings are readily achievable in integrated photonic systems through time-domain dynamic modulation. We experimentally study this concept using a two-resonator photonic molecule produced in a lithium niobate on insulator platform that is electro-optically modulated by rf stimuli. We demonstrate the dynamic tuning of the Hatano-Nelson coupling between the resonators, surpassing the asymmetry that has been achieved in previous work, to reach an EP for the first time. We are additionally able to flip the relative sign of the couplings for opposite directions by going past the EP. Using this capability, we show that the through-chain transport can be configured to exhibit both giant ($\sim 60$ dB) optical contrast as well as photonic gyration or non-reciprocal $\pi$ phase contrast.
\end{abstract}

\maketitle

In the analysis of photonic systems, couplings between two modes are routinely assumed to be symmetric, i.e., the same in both directions. Recent work on non-Hermitian dynamics that includes the effects of asymmetric couplings, and perhaps gain and loss, has demonstrated\cite{Gong_Ashida_Kawabata_Takasan_Higashikawa_Ueda_2018, Ashida_Gong_Ueda_2020, Wang_2021, Li_Wei_Cotrufo_Chen_Mann_Ni_Xu_Chen_Wang_Fan_2023} surprising phenomena like parity-time (PT) symmetry breaking \cite{Ruter_Makris_El-Ganainy_Christodoulides_Segev_Kip_2010, Regensburger_Bersch_Miri_Onishchukov_Christodoulides_Peschel_2012, Chang_Jiang_Hua_Yang_Wen_Jiang_Li_Wang_Xiao_2014, Peng_Ozdemir_Lei_Monifi_Gianfreda_Long_Fan_Nori_Bender_Yang_2014}, the non-Hermitian skin effect (NHSE) \cite{Hatano_Nelson_1996, Hatano_Nelson_1998, Okuma_Kawabata_Shiozaki_Sato_2020}, exceptional points (EPs) \cite{Graefe_Gunther_Korsch_Niederle_2008, Guo_Salamo_Duchesne_Morandotti_Volatier-Ravat_Aimez_Siviloglou_Christodoulides_2009, Ruter_Makris_El-Ganainy_Christodoulides_Segev_Kip_2010, Regensburger_Bersch_Miri_Onishchukov_Christodoulides_Peschel_2012,Demange_Graefe_2012, Peng_Ozdemir_Lei_Monifi_Gianfreda_Long_Fan_Nori_Bender_Yang_2014, Peng_Özdemir_Liertzer_Chen_Kramer_Yılmaz_Wiersig_Rotter_Yang_2016, Miri_Alu_2019}, and non-reciprocal transport \cite{Chang_Jiang_Hua_Yang_Wen_Jiang_Li_Wang_Xiao_2014, Thomas_Li_Ellis_Kottos_2016, Shao_Mao_Maity_Sinclair_Hu_Yang_Loncar_2020}. All of which lead to new physical insights and the potential for new applications.

\begin{figure*}[!t]
    \centering
    \includegraphics[width=1\linewidth]{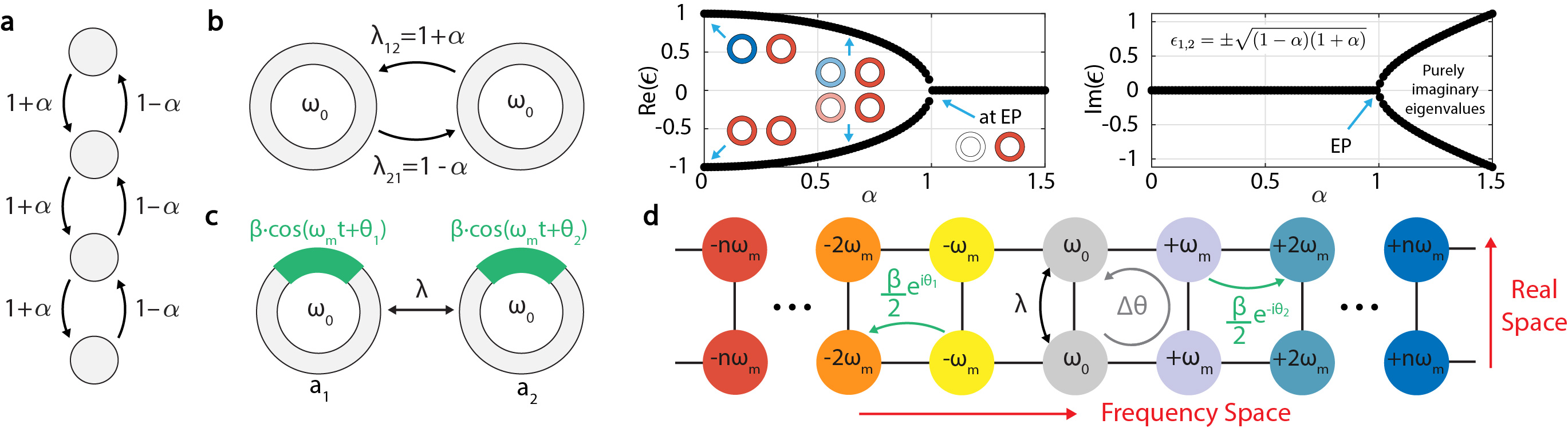}
    \caption{
        \textbf{(a)} Example Hatano-Nelson (HN) chain exhibiting non-reciprocal couplings for $\alpha \neq 0$. \textbf{(b)} A two resonator photonic molecule with imbalanced couplings $1 \pm \alpha$ has complex eigenvalues $\epsilon$ that vary with non-Hermitian perturbation $\alpha$. The exceptional point (EP) is indicated. The inset figures within the Re($\epsilon$) panel show evolution of the eigenmodes, with red and blue being opposite phases, and the color intensity signifying the amplitude.\textbf{(c)} Non-reciprocal coupling as in the HN model can be introduced to a photonic molecule with a \textit{reciprocal} inter-resonator coupling strength of $\lambda$ by means of time-domain modulation. The green regions represent the fraction of the photonic resonators that have time varying refractive index, which produces a temporal variance for the mode frequencies with a value of $\beta$. \textbf{(d)} We can generate a 2D \textit{time-invariant} equivalent representation of the time-varying photonic molecule from (c) by introducing a synthetic dimension in frequency space. The modulation depth $\beta$ sets the hopping rate along the synthetic dimension, while the relative phase $\Delta\theta$ sets the synthetic magnetic flux through each plaquette of the lattice.
    }
\label{fig:1}
\end{figure*}

One particularly important non-Hermitian system is the one-dimensional Hatano-Nelson chain \cite{Hatano_Nelson_1996,Hatano_Nelson_1998} in which the magnitudes of the couplings between adjacent elements are asymmetric (Fig.~\ref{fig:1}a), i.e., take different values depending on the coupling direction. This is the simplest system that gives rise to the NHSE and exhibits exceptional points. Hatano-Nelson type couplings, along with the NHSE and EPs, have been demonstrated in mechanical lattices \cite{Brandenbourger_Locsin_Lerner_Coulais_2019, Zhang_Yang_Ge_Guan_Chen_Yan_Chen_Xi_Li_Jia_2021} and electronic networks \cite{Helbig_Hofmann_Imhof_Abdelghany_Kiessling_Molenkamp_Lee_Szameit_Greiter_Thomale_2020} where non-reciprocal components, and therefore asymmetric couplings, are relatively easy to produce in real space. These capabilities are not available in photonics, however, so previous explorations of non-Hermitian physics have invoked synthetic dimensions \cite{Ozawa_Price_Goldman_Zilberberg_Carusotto_2016, Yuan:18} produced by dynamic modulations \cite{Weidemann_Kremer_2020, Wang_2021}, or through the explicit addition of optical gain \cite{Ruter_Makris_El-Ganainy_Christodoulides_Segev_Kip_2010, Regensburger_Bersch_Miri_Onishchukov_Christodoulides_Peschel_2012, Chang_Jiang_Hua_Yang_Wen_Jiang_Li_Wang_Xiao_2014, Peng_Ozdemir_Lei_Monifi_Gianfreda_Long_Fan_Nori_Bender_Yang_2014, Hodaei_Miri_Heinrich_Christodoulides_Khajavikhan_2014}. Specifically, Hatano-Nelson type couplings in photonics have only two extant real space demonstrations \cite{Liu_Wei_Hemmatyar_2022, Gao_et_al_PhysRevLett.130.263801} both of which used linking couplers having relatively large footprint, required active materials to produce the asymmetry and to overcome loss, and only achieved a modest asymmetry in the couplings.

In this work, we show that real space asymmetric couplings having very large contrast can be readily achieved in integrated photonics through the use of synthetic gauge fields \cite{Fang2012, Fang2013} induced by dynamic modulation. This approach was first demonstrated in microwave circuits \cite{Peterson:19} with good results, though an acousto-optics based photonic implementation \cite{Kim:2021te} produced only a weak effect. Here we leverage the very large electro-optic effect in thin film lithium niobate to produce the first Hatano-Nelson photonic system having coupling contrast large enough to reach, and even surpass, the EP condition. While past implementations of photonic Hatano-Nelson chains \cite{Weidemann_Kremer_2020,Wang_2021, Liu_Wei_Hemmatyar_2022} have focused on the NHSE eigenstates, our real space experimental system enables the study of transport through the chain, and the observation of non-reciprocal effects due to the persistent current found within the Hatano-Nelson system \cite{Zhang_et_al_PhysRevLett.125.126402}. With this capability, we are able to show that, at the EP, the system produces arbitrarily large optical contrast that may be beneficial for isolator and switch applications. By going past the EP, we are able to unidirectionally flip the sign of the coupling to achieve photonic gyration.
Let us start by considering the phenomenology of a simple photonic molecule composed of a pair of identical resonators having inter-resonator couplings $\lambda_{12} = 1+\alpha$ and $\lambda_{21}=1-\alpha$ (where $\alpha \in \mathbb{R}$) as shown in Fig.~\ref{fig:1}b. For any non-zero $\alpha$ the system can be analyzed as the Hatano-Nelson model and will exhibit the NHSE. Notably, an exceptional point (EP) can be found at the specific value $\alpha=1$ where the coupling for one direction becomes exactly zero \cite{Gong_Ashida_Kawabata_Takasan_Higashikawa_Ueda_2018, Miri_Alu_2019}. At this EP, both eigenmodes of the photonic molecule are fully localized onto a single resonator (in arbitrary length Hatano-Nelson chains the localization occurs on a single resonator at the end of the chain). Moreover, transport through the molecule (or chain) in the direction of the vanishing coupling gets shut down completely, producing an isolator-like response. For $\alpha > 1$, the energy spectrum of the photonic molecule changes from being real to purely imaginary as shown in Fig.~\ref{fig:1}b. Since the signs of the inter-resonator couplings $\lambda_{12}$ and $\lambda_{21}$ are opposite beyond the EP, transport through the molecule should exhibit a gyration response, i.e., a direction-dependent phase shift of $\pi$.
\begin{figure*}[t]
    \centering
    \includegraphics[width=1\linewidth]{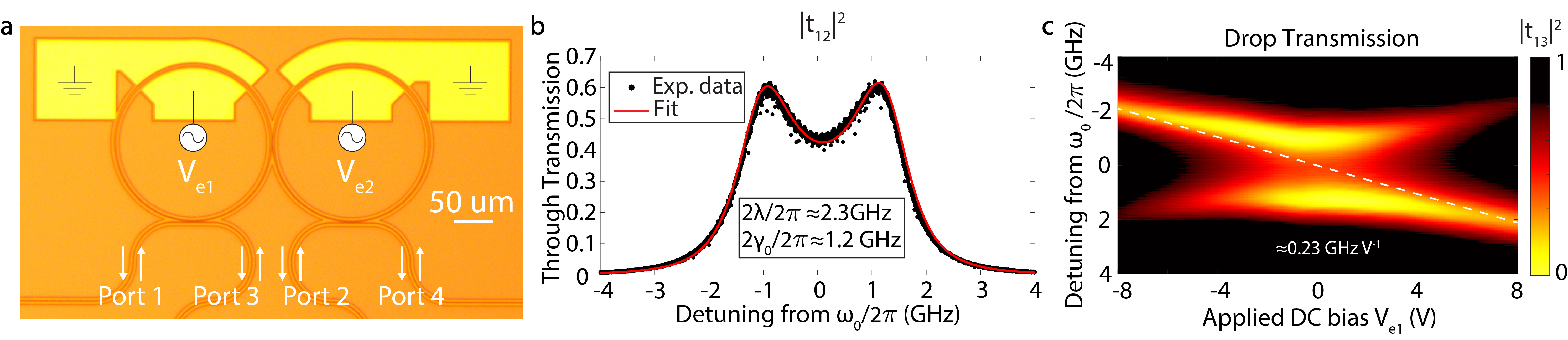}
    \caption{
        \textbf{(a)} True-color photograph of the experimental photonic molecule, fabricated in a thin film lithium niobate on insulator photonics platform. The rf modulation electrodes appear in brighter yellow. \textbf{(b)} The transmission spectrum $|t_{12}|^2$ measured through the unmodulated photonic molecule reveals the two optical supermodes of the device. \textbf{(c)} The left-port drop transmission spectrum $|t_{13}|^2$ measured as a function of dc bias electrode voltage $V_{e1}$ (where $V_{e2} = -V_{e1}$). This 2D visualization reveals an avoided crossing that allows us to extract the electro-optic modulation strength per applied voltage (i.e., $\beta/V$) through the slope of the white dashed line.
    }
\label{fig:2}
\end{figure*}
\vspace{7pt}
\\
\indent
Our implementation of Hatano-Nelson coupling through dynamic modulation is presented in Fig.~\ref{fig:1}c. We initially assume a reciprocal coupling $\lambda$ between the resonators. The resonators are then frequency modulated sinusoidally with identical frequency $\omega_m$, modulation depth $\beta$, and relative phase $\Delta\theta$, so that the instantaneous frequency of the $j^\textrm{th}$ resonator is expressed as $\omega_{j}=\omega_0+\beta \, \cos(\omega_m t+\theta_j)$ with $\theta_1 = 0$ and $\theta_2=\Delta\theta$. Setting aside loss, we can describe the time-dependent Hamiltonian for the system as $H(t)=\sum_{j}\hbar\left(\omega_0+\beta \, \cos(\omega_m t+\theta_{j})\right)\hat{a}_{j}^\dagger\hat{a}_{j}+\hbar\lambda(\hat{a}_{j}^\dagger\hat{a}_{j+1}+\hat{a}_{j}\hat{a}_{j+1}^\dagger)$ where $\hat{a}^\dagger$ $(\hat{a})$ is the creation (annihilation) operator. If we expand each operator as $\hat{a}_{j}=\sum_{n}{\hat{a}_{j,n}e^{i(\omega_0+n\omega_m) t}}$, where $n$ indexes the $n^\textrm{th}$ sideband, then we can apply the rotating frame approximation, to find:

\begin{align}
	H = & \sum_{j,n}\hbar (\omega_0+n\omega_m)\hat{a}_{j,n}^\dagger\hat{a}_{j,n} +\hbar\lambda\left(\hat{a}_{j,n}^\dagger\hat{a}_{j+1,n}+\hat{a}_{j,n}\hat{a}_{j+1,n}^\dagger\right) \notag \\
	& +\hbar\frac{\beta}{2}\left(\hat{a}_{j,n}^\dagger\hat{a}_{j,n+1}e^{-i\theta_j}+\hat{a}_{j,n}\hat{a}_{j,n+1}^\dagger e^{i\theta_j}\right) ~.
    \label{eq:ham}
 \end{align}

This Hamiltonian describes a two-dimensional lattice (Fig.~\ref{fig:1}d) with one dimension $j$ representing space and the other dimension $n$ corresponding to a frequency domain synthetic dimension \cite{Peterson:19,Kim:2021te}. Here, the synthetic dimension is invoked to understand the effect of dynamic modulation, which leads to a gauge field and produces an effective asymmetric coupling in real space. The first term of {Eq.~(\ref{eq:ham})} is analogous to the an electric potential of a uniform electric field in the synthetic dimension {\cite{Yuan:16}}. The latter terms resemble the Hamiltonian of a charged particle moving on this lattice subjected to a uniform magnetic field~\cite{Tzuang_Fang_Nussenzveig_Fan_Lipson_2014}, having a flux set by the relative modulation phase $\Delta\theta$. At this point we can discard the synthetic dimension and generate a new effective Hamiltonian for the central chain around $\omega_0$ (details in Supplement $\S1$) by projecting the lattice sites for non-zero sidebands (i.e., non-zero potential) into effective coupling pathways for the chain of the zeroth sideband. With this new perspective, we find that the effective Hamiltonian for the zeroth subband matches the Hatano-Nelson model. It is also useful to note that the application of other modulation frequencies that are incommensurate with $\omega_m$ will produce orthogonal synthetic spaces, and therefore can be used to generate pairwise non-reciprocal couplings between other resonators in more complex configurations. 

While the exact coupling rates can be evaluated numerically, a simpler treatment with only the first-order upper and lower sidebands can be resolved into an analytical expression that helps to generate intuitive understanding. In this simplified treatment, the coupling between the sites at zero potential ($\omega_0$) occurs only via three paths: through \textit{reciprocal} hopping $\lambda$, and through the upper and lower frequency shifted channels that undergo non-reciprocal phase shift due to the synthetic magnetic flux. The net coupling at $\omega_0$ is therefore affected by interference between these three channels, and for the ideal non-reciprocal case with $\omega_{m}=\lambda$ and $\Delta\theta=\pi/2$ {\cite{Peterson:19}}, is evaluated as
\begin{equation}
	\lambda_{12,21}=\lambda\left(1\pm\frac{\lambda \beta^2}{(4\lambda^2+\gamma_0^2)\gamma_0}\right)  ~.
 \label{eq:eff_lambda}
\end{equation}
Here, we have also included $\gamma_0$, which is half of the loss rate of each resonator within the photonic molecule in the absence of temporal modulation. From Eq.~(\ref{eq:eff_lambda}), we can easily see that the effective coupling takes the form $\lambda(1 \pm \alpha)$ similar to the Hatano-Nelson model~\cite{Hatano_Nelson_1996,Hatano_Nelson_1997,Hatano_Nelson_1998} with $\alpha=\lambda \beta^2/(4\lambda^2+\gamma_0^2)\gamma_0$. Furthermore, the effective loss rate for each site of the central chain also exhibits a modulation-dependent behavior, as elaborated in Supplement Section $\S1$. 

Our experimental implementation was performed at 1550 nm using a 500 nm thick X-cut LiNbO\textsubscript{3}-on-insulator integrated photonics platform. The large electro-optic coefficients of LiNbO\textsubscript{3} (LN) enable rf-induced frequency modulation of the integrated microring resonators. The resonators within the photonic molecule (Fig.~\ref{fig:2}a) support TE\textsubscript{00} modes, and were designed with a crescent-like ring geometry  \cite{Liang_Huang_Mohanty_Shin_Ji_Carter_Shrestha_Lipson_Yu_2021} that was produced by introducing a 1.125 um shift between two circles of diameter of 195.25 um and 200 um. The crescent-like shape facilitates high-Q devices with a tight bending radius, allowing us to achieve overcoupling (where extrinsic coupling rate exceeds the intrinsic loss rate) at the input and output waveguides without increasing the intrinsic loss rate, which enables a high transmission coefficient. Each resonator is accessed through a single-mode waveguide with grating couplers on both ends to provide off-chip optical access. Metal electrodes for inducing modulation are fabricated through e-beam evaporation of 300 nm gold followed by a lift-off process. The electrodes were designed so that the applied rf field is mostly oriented along the Z-axis of the LN crystal to harness the r\textsubscript{33} electro-optic coefficient. To achieve a balance where we have large modulation depth with low optical loss, the gap between the electrode and the resonator was set to 2.4 microns, and the angular coverage of the modulation electrode (green region in Fig.~{\ref{fig:1}}c) was designed as $\pm 45$ degrees with respect to the Z axis of the crystal.

\begin{figure*}[th!]
    \centering
    \includegraphics[width=1\linewidth]{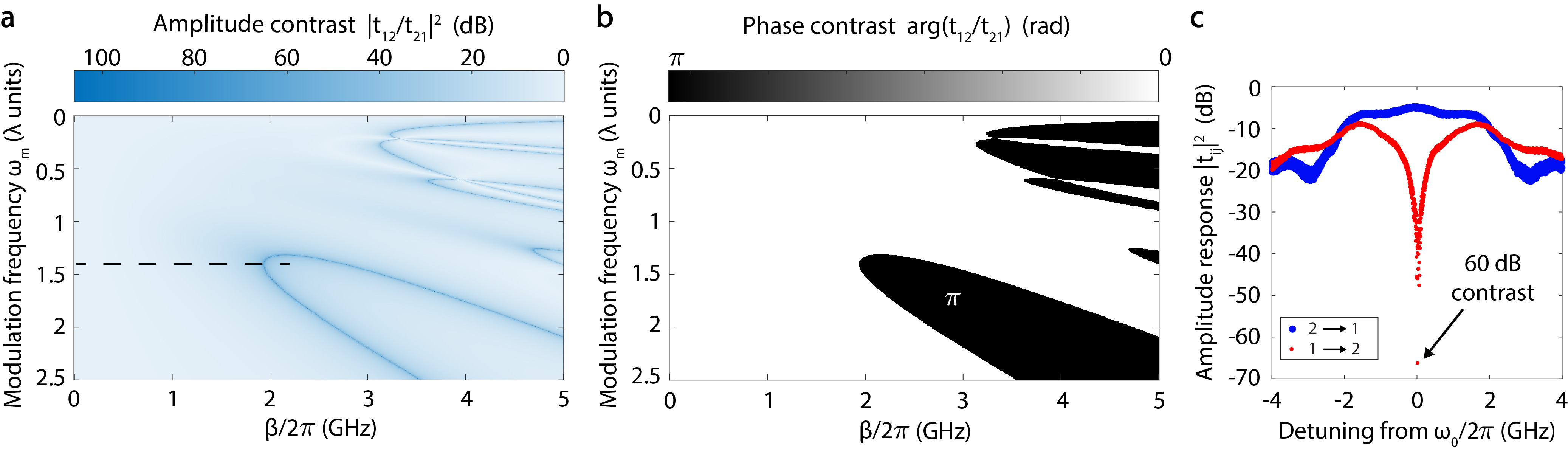}
    \caption{
        Numerical simulation of the anticipated optical contrast and experimental demonstration of giant non-reciprocity. (\textbf{a}) Simulated amplitude contrast $\left| t_{12}/t_{21} \right|^2$ and (\textbf{b}) phase contrast $\textrm{arg}(t_{12}/t_{21})$ at zero detuning, and using experimentally extracted parameters. The dashed line corresponds to a parameter sweep that we perform in experiments shown in Fig.~\ref{fig:4}. \textbf{(c)} Experimental transmission spectrum through the photonic molecule for the forward and backward directions at 1.6 GHz with an applied voltage of the rf control signal $V_{ec} \approx 6.77$ V$_{rms}$. 
    }
\label{fig:3}
\end{figure*}

In order to measure optical transport through the chain we make use of a heterodyne detection system (Supplement $\S2$). Here we define the optical transfer function between any pair of ports $p$ and $q$ as the field ratio $t_{pq} = s_{p,\textrm{-}}/s_{q,\textrm{+}}$, where $s_{p,\textrm{-}}$ is the outgoing and $s_{q,\textrm{+}}$ is the incoming wave. Figure~\ref{fig:2}b presents the power transmission spectrum through the resonator chain without any modulation applied (i.e. $|t_{21}|^2 = |t_{12}|^2$), revealing two distinct peaks corresponding to the symmetric and antisymmetric supermodes of the system. This enables us to estimate the inter-resonator coupling $\lambda/2\pi \approx 1.15$ GHz. Similarly, the loaded optical loss rates are estimated as $2\gamma_0/2\pi \approx 1.2$ GHz corresponding to a Q factor of 167,000. Electro-optic modulation is then tested by applying a dc bias voltage to the electrodes in a pull-push configuration ($V_{e2} = - V_{e1}$), shifting the resonance frequencies of the rings in opposite directions. For a transmission measurement (Fig.~\ref{fig:2}c) performed at the left side `drop' port (i.e. $|t_{13}|^2$) we observe an avoided crossing as a function of $V_{e1}$, which allows the estimation of the electro-optic modulation strength $\beta$ from the tuning slope. The specific value of $\beta$ is determined by both the electric field strength (i.e., gap between the electrodes), and the angular coverage of the electrodes for the ring. For our configuration we extract $\beta/(2\pi\times V_{ec} )\approx 0.23$ GHz/V. 

We can now investigate the parameter space by extrapolating from the measured optical properties and electro-optic modulation strength. In Fig.~\ref{fig:3}a we present the numerically simulated ratio of forward and backward power transmission $\left| t_{12}/t_{21} \right|^2$, i.e., the amplitude contrast, as well as the phase contrast $\textrm{arg}(t_{12}/t_{21})$ at the zero detuning frequency $\omega_0$. For both plots we vary the synthetic lattice spacing (modulation frequency) $\omega_m$ and hopping strength $\beta$, while setting $\Delta\theta = \pi/2$ to maximize the optical non-reciprocity. In Fig.~\ref{fig:3}a we observe many trajectories in the parameter space where very high optical contrast is obtained, corresponding to EPs in the photonic molecule. When $\beta$ is increased past an EP trajectory, the sign of the inter-resonator coupling flips, resulting in an island-like region where gyration takes place, i.e., a direction sensitive $\pi$-phase contrast. These structures in parameter space are fairly complex since we employed 5 sidebands in the calculation (we found additional sidebands beyond the 5 we used do not noticeably affect the simulation). Importantly, Fig.~\ref{fig:3} predicts that the most accessible EP with the lowest value of $\beta$ will be found near $\omega_m \approx 1.4\lambda$. 

\begin{figure*}[!t]
    \centering
    \includegraphics[width=1\linewidth]{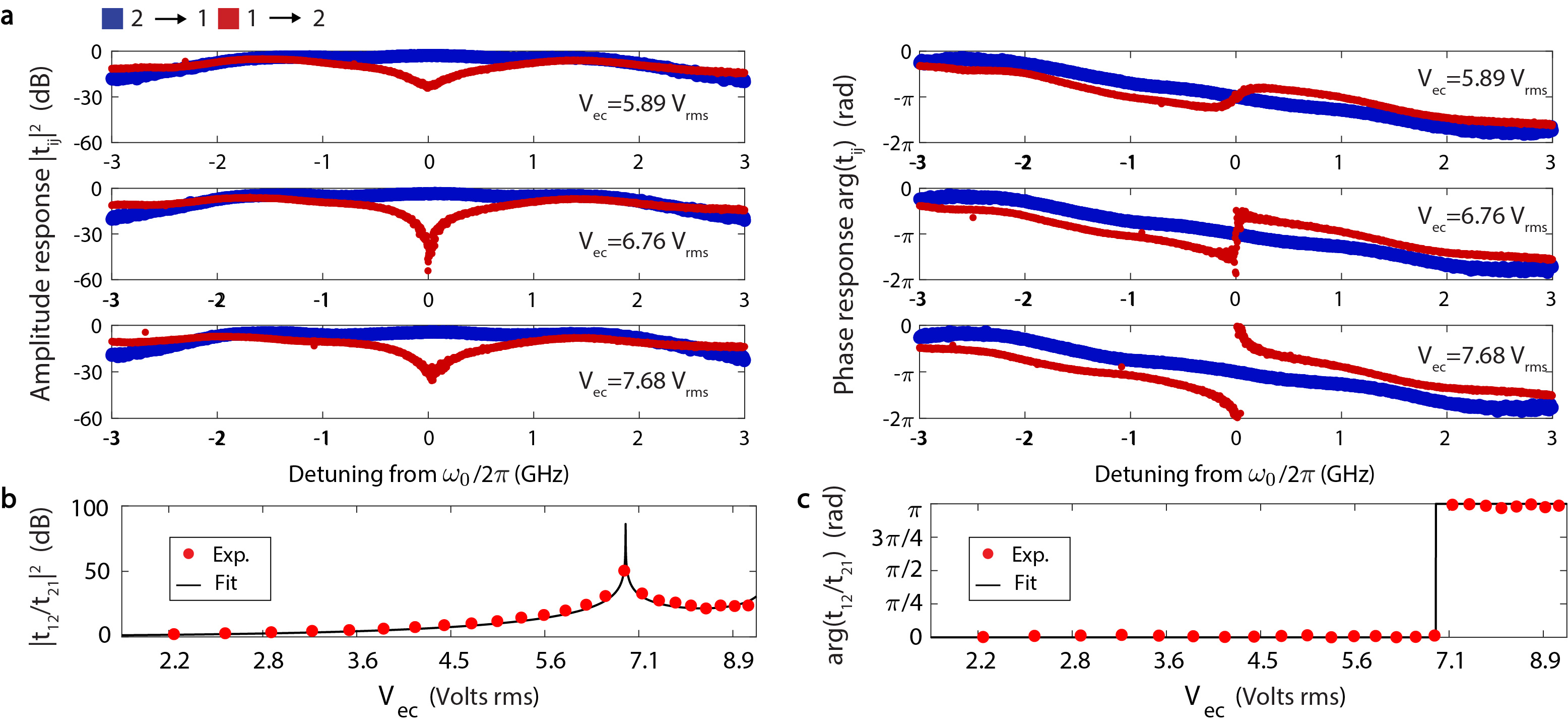}
    \caption{
	\textbf{(a)} Experimentally measured representative spectra of the amplitude and phase responses before, near, and after the EP at $V_{ec, crit} \approx 6.77$ volts rms. Experimentally measured evolution of the non-reciprocal \textbf{(b)} amplitude contrast $\left| t_{12}/t_{21} \right|^2$  and \textbf{(c)} phase contrast $\textrm{arg}(t_{12}/t_{21})$ at zero detuning. The sweep of $V_{ec}$ corresponds to $\beta$ values roughly along the dashed line shown in Fig.~\ref{fig:3}a. These measurements confirm the maximal transmission contrast at the EP and gyration beyond the EP.
    }
\label{fig:4}
\end{figure*}

For experiments we generate an rf signal with frequency $\omega_m$. The amplitude of modulation  $V_{ec}$ (rms value) controls the hopping strength $\beta$ in the synthetic dimension. Specifically, we set $V_{e1} = \sqrt{2} \, V_{ec} \cos(\omega_m t)$ and $V_{e2} = \sqrt{2} \, V_{ec} \cos(\omega_m t + \Delta\theta)$ with the phase offset of $\Delta\theta = \pi/2$. Figure~\ref{fig:3}c shows that we successfully reached the above predicted EP for $V_{ec, crit} \approx 6.77$ volts rms and were able to experimentally confirm a record amplitude contrast of $\approx 60$ dB. Our ability to measure and report the optical isolation contrast was limited only by the noise floor of the measurement electronics, though in principle the contrast should tend to infinity at the EP with sufficient fine tuning and improved SNR.

To fully extract the amplitude and phase information of the transfer functions, we modify our setup slightly to include an external electro-optic amplitude modulator (details in Supplement $\S3$). We are specifically interested in the evolution of the optical transfer functions $t_{21}$ and $t_{12}$, i.e., in opposite directions, as we sweep past the EP along the path shown by the dashed line in Fig.~\ref{fig:3}a. Figure~\ref{fig:4}a shows a representative sequence of the evolution of these transfer function spectra (both amplitude and phase) for $V_{ec}$ just prior to, near, and after $V_{ec, crit}$. Figure~\ref{fig:4}b,c present consolidated single-point (zero detuning) measurements of the experimental transmission for many more values of $V_{ec}$ through the identified sweeping path. From here we can verify that the non-reciprocal contrast has a peak around $V_{ec,crit}$, in agreement with Fig.~\ref{fig:3}c. Additionally, we confirm the direction dependent $\pi$-phase contrast for $V_{ec}$ beyond the critical value, while the amplitude contrast also reduces, demonstrating that we have entered the predicted gyration island. Notably, at zero detuning, the slope of the phase response indicates a fast light (slow light) dispersion for backward propagation direction before (after) the critical voltage level. The slope of the phase response becomes remarkably steep, eventually approaching infinity at $V_{ec,crit}$. This behaviour also serves as an indication of the EP, which is realized by destructive interference between the channels of the different sidebands. 

The synthesis of asymmetric, non-Hermitian couplings in real space is important for the exploration of a variety of novel physical systems, including topological materials~\cite{Ezawa_2019,Lee_Li_Gong_2019,Helbig_Hofmann_Imhof_Abdelghany_Kiessling_Molenkamp_Lee_Szameit_Greiter_Thomale_2020,Borgnia_Kruchkov_Slager_2020,Zhang_Yang_Fang_2020,Okuma_Kawabata_Shiozaki_Sato_2020,Ou_Wang_Li_2023}, high-order exceptional points \cite{Graefe_Gunther_Korsch_Niederle_2008,Demange_Graefe_2012, Miri_Alu_2019}, and skin effect phenomena~\cite{Hatano_Nelson_1996, Hatano_Nelson_1998, Brandenbourger_Locsin_Lerner_Coulais_2019, Hofmann_Helbig_Lee_Greiter_Thomale_2019,Helbig_Hofmann_Imhof_Abdelghany_Kiessling_Molenkamp_Lee_Szameit_Greiter_Thomale_2020, Okuma_Kawabata_Shiozaki_Sato_2020, Helbig_Hofmann_Imhof_Abdelghany_Kiessling_Molenkamp_Lee_Szameit_Greiter_Thomale_2020, Zhang_Yang_Ge_Guan_Chen_Yan_Chen_Xi_Li_Jia_2021, Liu_Wei_Hemmatyar_2022, Gao_et_al_PhysRevLett.130.263801}. Our work demonstrates a versatile approach to achieve asymmetric couplings in real space using commonly available modulation techniques that can be implemented in photonics foundries without requiring optical gain. Moreover, this approach is shown to be very effective in reaching the EP condition, and surpassing it to produce gyration, neither of which were achieved in previous integrated photonics efforts. Our technique can be readily expanded to more complex configurations by using multiple modulation frequencies that produce orthogonal synthetic spaces, enabling higher dimensional photonic materials and longer-range asymmetric couplings.

\begin{acknowledgments}
This work was funded by the US Office of Naval Research Multi-University Research Initiative (MURI) under grant N00014-20-1-2325, the US Air Force Research Lab under award FA9453-20-2-0001, and the Defense Advanced Research Projects Agency (DARPA) under cooperative agreement D24AC00003. R.T. acknowledges funding by the Deutsche Forschungsgemeinschaft (DFG, German Research Foundation) through Project-ID 258499086 - SFB 1170 and through the W{\"u}rzburg-Dresden Cluster of Excellence on Complexity and Topology in Quantum Matter – ct.qmat Project-ID 390858490 - EXC 2147. J. N. acknowledges support from the U.S. Department of Energy (DOE), Office of Basic Energy Sciences, Division of Materials Sciences and Engineering and performed, in part, at the Center for Integrated Nanotechnologies, an Office of Science User Facility operated for the U.S. DOE Office of Science. Sandia National Laboratories is a multimission laboratory managed and operated by National Technology and Engineering Solutions of Sandia, LLC., a wholly owned subsidiary of Honeywell International, Inc., for the U.S. DOE’s National Nuclear Security Administration under contract DE-NA0003525. The views expressed in the article do not necessarily represent the views of the U.S. DOE or the United States Government. P.Z. was primarily supported by the Center for Emergent Materials, an NSF MRSEC, under award number DMR-2011876. The authors acknowledge helpful discussions with Violet Workman and Zhiyin Tu.
\end{acknowledgments}

\setcounter{figure}{0}
\renewcommand{\figurename}{Fig.}
\renewcommand{\thefigure}{S\arabic{figure}}
\setcounter{equation}{0}
\renewcommand{\theequation}{S\arabic{equation}}

\section{Supplementary information: Giant non-reciprocity and gyration through modulation-induced Hatano-Nelson coupling in integrated photonics}

\section{Input-output relation of photonic molecule under dynamic modulation}

We use coupled mode theory to calculate the input-output relationships for the photonic molecule considered in this work. Each resonator has an intrinsic resonance frequency $\omega_{0}$, and the \textit{reciprocal} coupling rate between resonators is $\lambda$. We sinusoidally modulate the resonance at each site with a modulation frequency of $\omega_{m}$. The phase of the modulation is changed linearly with fixed phase offset of $\Delta\theta$ relative to the adjacent sites. Generally we can write that the resonance frequency of the $n$th resonator is:
\begin{equation}
    \omega_{n}(t) = \omega_{0} + \beta \cos(\omega_{m}t+\theta_n)
    \label{eq:DynTimeDomain}
\end{equation}
where $\beta$ is the modulation amplitude. We can also express the time evolution of the intracavity field in each resonator as~\cite{Peterson:19,Kim:2021te}:
\begin{equation}
	\frac{\partial}{\partial t}\ket{a(t)}=\left[i\Omega_{0} + i\Omega_{1}(t)-\Gamma \right]\ket{a(t)} + iK^{T}\ket{s_{+}(t)}
	\label{eq:TemporalCMT}
\end{equation}
Here, $\ket{a(t)}=\left[a_{1}(t), a_{2}(t)\right]^{T}$ is a vector containing the intracavity field of each resonator, and the other variables are: 
\begin{multline}
	\Omega_0=\begin{bmatrix}
		\omega_{0} & \lambda  \\
		\lambda & \omega_{0}  \\
	\end{bmatrix},
 	 \Gamma=\begin{bmatrix}
		\gamma_0 & 0  \\
		0 & \gamma_0  \\
	\end{bmatrix}, 	 
        K=\begin{bmatrix}
		\sqrt{k_{ext}} & 0  \\
		0 & \sqrt{k_{ext}}  \\
	\end{bmatrix},\\
	\Omega_1=\begin{bmatrix}
		\beta\cos(\omega_{m} t+\theta_1) & 0  \\
		0 & \beta\cos(\omega_{m} t+\theta_2)  \\
	\end{bmatrix}
	\label{eq:Omega_0}
\end{multline}

\begin{figure}[t!]
    \includegraphics[width=1\linewidth]{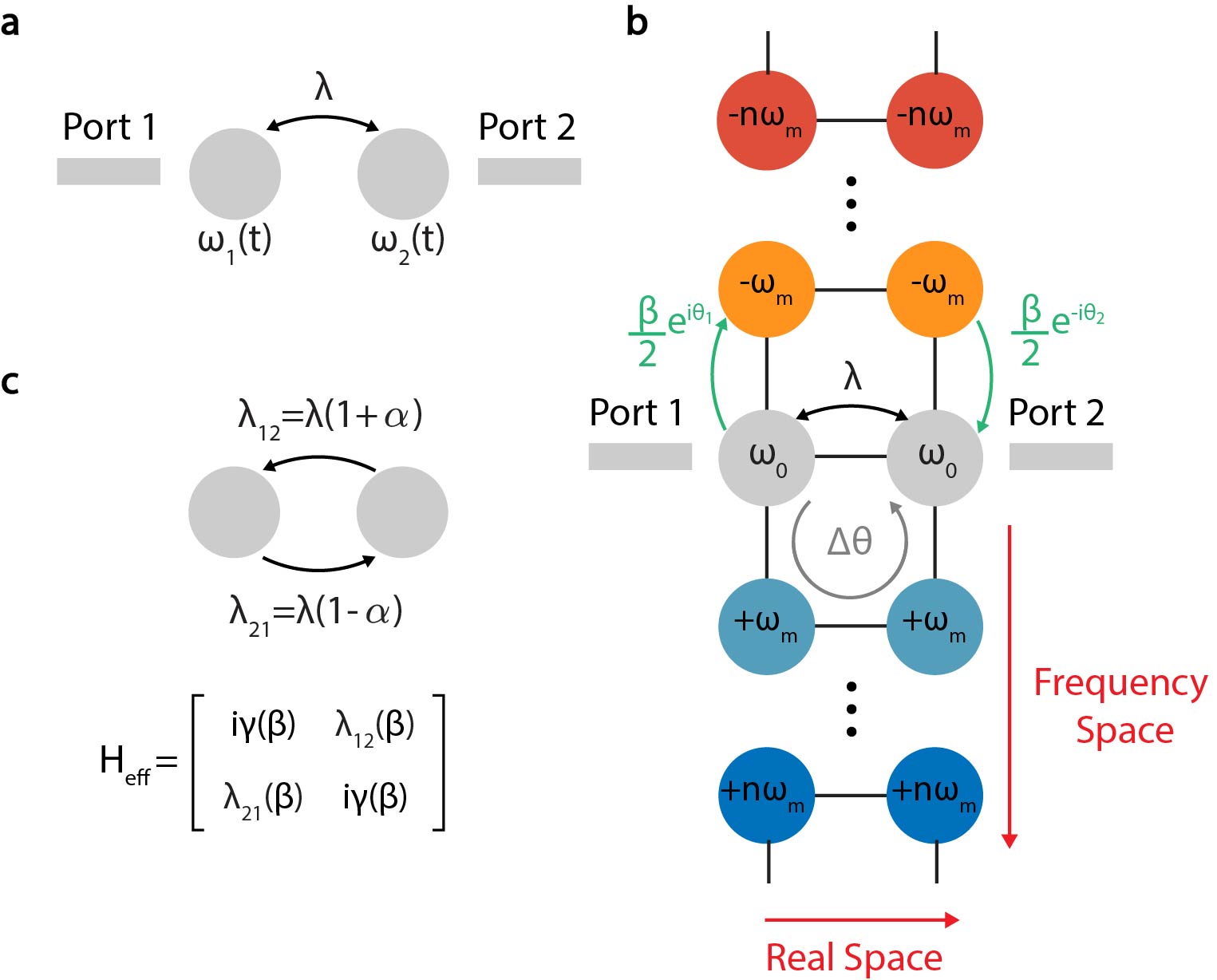}
    \centering
    \caption{\textbf{(a)} The configuration consists of 2 coupled resonators. Each resonator is assigned a time-varying resonant frequency, as described by Eqn.~\ref{eq:DynTimeDomain}. \textbf{(b)} This short chain can be mapped to a time-invariant 2D lattice, with synthetic electrical and magnetic fields. \textbf{(c)} All the interactions taking place in the 2D space can then be collapsed into an effective system for the central resonator pair and reveals a Hatano-Nelson type non-reciprocal coupling.
    }
    \label{supfig:1}
\end{figure}

The effect of the modulation is taken into account via $\Omega_{1}(t)$ matrix, and $\Gamma$ includes on-diagonal elements $\gamma_0$, which are the half of the total decay rate of each resonator. $k_{ext}$ in $K$ is the external coupling rate between the resonators and the ports, and $\Gamma$ and $K$ are related to each other as $2\Gamma=K^{\dagger}K+\kappa I$~\cite{IEEEJQE40Suh}, where $\kappa=\textrm{diag}(\kappa_0, \kappa_0)$ is the intrinsic loss in each resonator. Finally, the incoming and outgoing waves from the ports are expressed by $\ket{s_{\pm}(t)}$ vectors respectively. For an input excitation of $\ket{s_{+}(t)}$, we can write the outgoing wave as:
\begin{equation}
	\ket{s_{-}(t)} = \ket{s_{+}(t)} + iK\ket{a(t)}
    \label{eq:InputOutput}
\end{equation}

In order to calculate the spectral response, we take the Fourier transform (i.e. $\ket{a(\omega)} = \int dt\ket{a(t)}e^{-i\omega t}$) of both sides of Eq.~\ref{eq:TemporalCMT}:
\begin{align}
	i\omega \ket{a(\omega)} = & \, iH_{0}\ket{a(\omega)} + iK^{T}\ket{s_{+}(\omega)} \nonumber \\
	&+B\ket{a(\omega-\omega_{m})}-B^{\dagger}\ket{a(\omega+\omega_{m})},
	\label{eq:DynFreqDomain}
\end{align}
where $H_{0}$ is the Hamiltonian of the system without any modulation and is given by $H_{0}=\Omega_{0}+i\Gamma$. The matrix $B=\textrm{diag}(\beta_{1},\beta_{2})$ represents the coupling between the adjacent resonators in the synthetic frequency space with $\beta_{n} =e^{-i\theta_{n}}\beta/2 $. The eqn.~\ref{eq:DynFreqDomain} produces a recursive relationship between different frequency components separated by $\pm \omega_m$. Since we generate infinitely many sidebands due to the applied modulation, we can extend the Eq.~\ref{eq:DynFreqDomain} for all sidebands as~\cite{Peterson:19}:
\begin{equation}
	i\omega\ket{\alpha(\omega)} = \mathcal{H}\ket{\alpha(\omega)} + i\mathcal{K}^{T}\ket{\sigma_{+}(\omega)},
    \label{eq:2DArrayEq}
\end{equation}
where $\mathcal{H}$ is the block-tridiagonal matrix
\begin{equation}
	\mathcal{H}=\begin{bmatrix}
		\ddots & \ddots & 0 & 0 & 0 \\
		\ddots & iH_{0}-i\omega_{m}I & B & 0 & 0 \\
		0 & -B^{\dagger} & iH_{0} & B & 0 \\
		0 & 0 & -B^{\dagger} & iH_{0}+i\omega_{m}I & \ddots \\
		0 & 0 & 0 & \ddots & \ddots	
	\end{bmatrix},
	\label{eq:FreqSpaceHam}
\end{equation}
and $\mathcal{K}$ is a block-diagonal matrix with each block component being $K$. Furthermore, the intracavity field vector for all sidebands $\ket{\alpha(\omega)}$ and incoming and outgoing wave vectors $\ket{\sigma_{\pm}(\omega)}$ are:
\begin{equation}
	\ket{\alpha(\omega)} =
	\begin{bmatrix}
		\vdots \\
		\ket{a(\omega+\omega_{m})}\\
		\ket{a(\omega)}\\
		\ket{a(\omega-\omega_{m})}\\
		\vdots
	\end{bmatrix},\,\,
	\ket{\sigma_{\pm}(\omega)} =
	\begin{bmatrix}
		\vdots \\
		\ket{s_{\pm}(\omega+\omega_{m})}\\
		\ket{s_{\pm}(\omega)}\\
		\ket{s_{\pm}(\omega-\omega_{m})}\\
		\vdots
	\end{bmatrix}.
\end{equation}
We see that once eqn.~\ref{eq:DynFreqDomain} is re-written as  eqn.~\ref{eq:2DArrayEq} with a Hamiltonian of eqn.~\ref{eq:FreqSpaceHam}, our system now transforms into a 2D lattice (Fig.~\ref{supfig:1}b) spanning in frequency and spatial domain. Furthermore, the Hamiltonian in eqn. \ref{eq:FreqSpaceHam} for the two-dimensional array not only spans in the synthetic frequency dimension, but also manifests two synthetic fields in the system. The diagonal terms of the Hamiltonian are equally spaced by $\omega_{m}$, which depends on the external modulation frequency. This is equivalent to the Hamiltonian for a charged particle in an electric field that produces an electrostatic potential gradient along the direction of the field. Moreover, examining $B$ matrix, which represents the coupling between the adjacent resonators in the synthetic frequency space, shows that there exists a direction-dependent phase shift of $\Delta\theta$ in each plaquette, induced by the phase gradient $\Delta\theta$ with the external modulation (Eq. \ref{eq:DynTimeDomain}). This is equivalent to the Hamiltonian for a charged particle in a magnetic field with out of plane flux of $\Delta\theta$ \cite{PhysRevLett108Fang,Tzuang_Fang_Nussenzveig_Fan_Lipson_2014}.
\begin{figure*}[ht!]
    \includegraphics[width=1\linewidth]{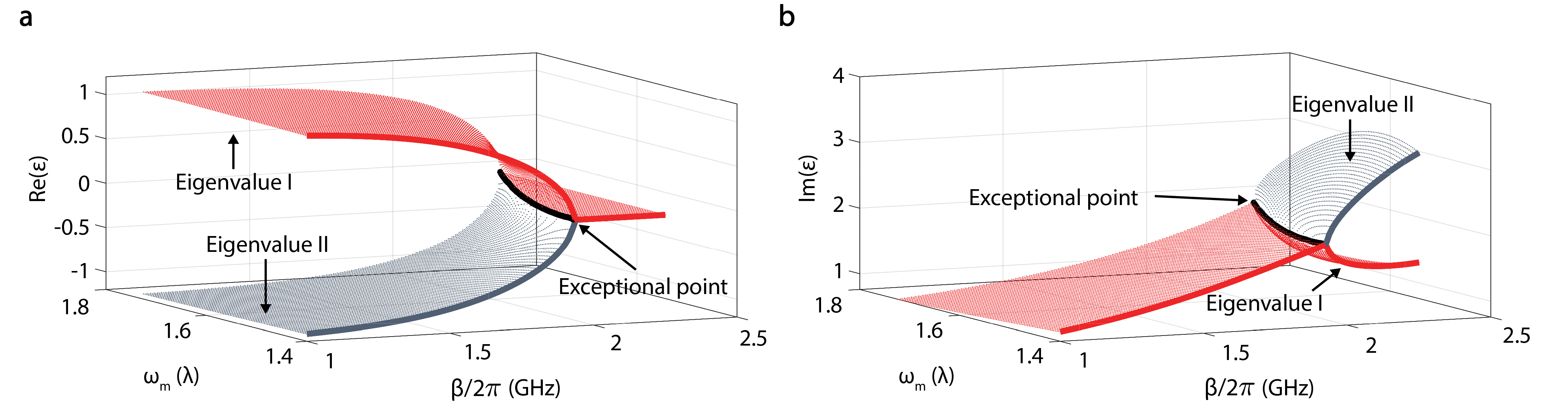}
    \centering
    \caption{
    Simulated eigenvalues of the photonic molecule for various modulation depth ($\beta$) and frequency ($\omega_m$) by including five sidebands above and below $\omega_0$. The real \textbf{(a)} and imaginary \textbf{(b)} solutions coalesce to a single point at the exceptional points, which are highlighted as a black curve in this parameter space.
    }
    \label{supfig:4}
\end{figure*}
The impact of the synthetic electric and magnetic fields on the transport properties of the photonic molecule can be determined by producing an effective Hamiltonian for the central chain. To do so, we project the effect of sites at non-zero potential by using the properties of the block tridiagonal matrix as:
\begin{equation}
    i\omega I\ket{a(\omega)}=\left\{i[H_{0}]-O-\Tilde{O} \right\}\ket{a(\omega)}+iK^{T}\ket{s_{+}(\omega)}
    \label{eq:projection}
\end{equation}
where 
\begin{align}
    O&=B([i\omega I-iH_{-1}] +B([i\omega I-iH_{-2}]+...\\ \nonumber
    &+B([i\omega I-iH_{-J+1}]+B[i\omega I-iH_{-J}]^{-1}
    (B)^{\dagger})^{-1}\\ \nonumber
    &(B)^{\dagger}...)^{-1})(B)^{\dagger}
\end{align}
and 
\begin{align}
    \Tilde{O}&=(B)^{\dagger}([i\omega I-iH_{1}] +(B)^{\dagger}([i\omega I-iH_{2}]+...\\ \nonumber
    &+(B)^{\dagger}([i\omega I-iH_{J-1}]+(B)^{\dagger}[i\omega I-iH_{J}]^{-1}B)^{-1}\\
    \nonumber
    &B...)^{-1})B
\end{align}
Here, we truncated the synthetic frequency dimension at $\pm J^{\textrm{th}}$ sidebands. Furthermore, $H_{j}\equiv H_{0}-j\omega_{m}I$, $\ket{a_{j}}\equiv \ket{a(\omega+j\omega_{m})}$ and $\ket{s_{-,j}}\equiv \ket{s_{-}(\omega+j\omega_{m})}$, where $\{j\in \mathbb{Z}|-J\leq j\leq J\}$. From eqn.~\ref{eq:projection}, we can also write the effective Hamiltonian for the central chain around $\omega_0$ as:
\begin{equation}
   iH_\textrm{eff}=i[H_{0}]-O-\Tilde{O}=\begin{bmatrix}
 -\gamma & i\lambda_{12}\\
 i\lambda_{21} & -\gamma
 \end{bmatrix}
    \label{eq:effective_Hamiltonian}
\end{equation}
Here, $\gamma$ is the effective loss rate, $\lambda_{12}$ and $\lambda_{21}$ are the directional coupling rates of the sites at zero potential with $\lambda_{12}\neq\lambda_{21}$. Also, $\gamma$, $\lambda_{12}$ and $\lambda_{21}$ are purely real due to the symmetric (asymmetric) transmission (phase) spectrum of the photonic molecule around $\omega_0$. Because of this symmetry, the effective Hamiltonian is pseudo anti-Hermitian and shows an exceptional point (EP) when $\lambda_{12}$ or $\lambda_{21}$ is zero. To understand the formation of the EP, we obtain a simplified expression of $H_\textrm{eff}$ by considering the sidebands up to first order as:
\begin{align}
   H_\textrm{eff}&=\begin{bmatrix}
 i\gamma_0 & \lambda\\
 \lambda & i\gamma_0
 \end{bmatrix}\nonumber\\
 &+
 \begin{bmatrix}
 \omega_m-i\gamma_0 & +\lambda e^{-i\Delta\theta}\\
 +\lambda e^{i\Delta\theta} & \omega_m-i\gamma_0
 \end{bmatrix}\frac{\beta^2/4}{(\omega_m-i\gamma_0)^2-\lambda^2} \nonumber \\
  &+\begin{bmatrix}
 -\omega_m-i\gamma_0 & +\lambda e^{i\Delta\theta}\\
 +\lambda e^{-i\Delta\theta} & -\omega_m-i\gamma_0
 \end{bmatrix}\frac{\beta^2/4}{(\omega_m+i\gamma_0)^2-\lambda^2}
    \label{eq:effective_Hamiltonian}
\end{align}

Considering the symmetry of the photonic molecule, we can further simplify the above expression by substituting $\omega_m=\lambda$ and $\Delta\theta=\pi/2$:
\begin{align}
 \gamma&=\gamma_0\left(1+\frac{\beta^2(\gamma_0^2+2\lambda^2)}{2(\gamma_0^4+4\gamma_0^2\lambda^2)}\right), \nonumber \\ 
 \lambda_{12}&=\lambda\left(1+\frac{\beta^2\lambda\gamma_0}{\gamma_0^4+4\gamma_0^2\lambda^2}\right), \nonumber \\\lambda_{21}&=\lambda\left(1- \frac{\beta^2\lambda\gamma_0}{\gamma_0^4+4\gamma_0^2\lambda^2}\right)
 \label{eq:eff_param}
\end{align}

From eqn.~\ref{eq:eff_param}, we see that effective loss rate ($\gamma$) and inter resonator coupling rates ($\lambda_{12}$ and $\lambda_{21}$) show a modulation dependent behaviour. Importantly, the inter resonator coupling rates are non-reciprocal (i.e., $\lambda_{12}\neq\lambda_{21}$ and $\lambda_{12,21}=\lambda(1 \pm \alpha)$), and show a dependence similar to the Hatano-Nelson model with $\alpha=\beta^2\lambda\gamma_0/(\gamma_0^4+4\gamma_0^2\lambda^2)$. We reach the EP when $\alpha$ becomes 1, resulting in infinite non-reciprocal contrast for transmission through the photonic molecule. If we include higher order sidebands, the same form of the effective Hamiltonian holds, and we show this by numerically calculating the eigenvalues of the system in Fig.~\ref{supfig:4}. Similar to previous conclusion, we see that a set of exceptional points occurs for various values of $\beta$ and $\omega_m$ (shown as the black curve). Here, the bold red and grey curves in Fig.~\ref{supfig:4} correspond to our experimental results that are presented in the main manuscript.

After obtaining the effective Hamiltonian, we can also calculate the S parameters of the photonic molecule with two external ports as:
\begin{equation}
    S_{0}=I+K[{\omega I-H_\textrm{eff}}]^{-1}K^{T}.
    \label{eq:RealSMatrix}
\end{equation}
In Fig.~3 of the main manuscript we present the implications of Eq.~\ref{eq:RealSMatrix} in the form of a heat map showing the non-reciprocity in the amplitude and phase response through the photonic molecule.

\begin{figure}[h!]
    \includegraphics[width=1\linewidth]{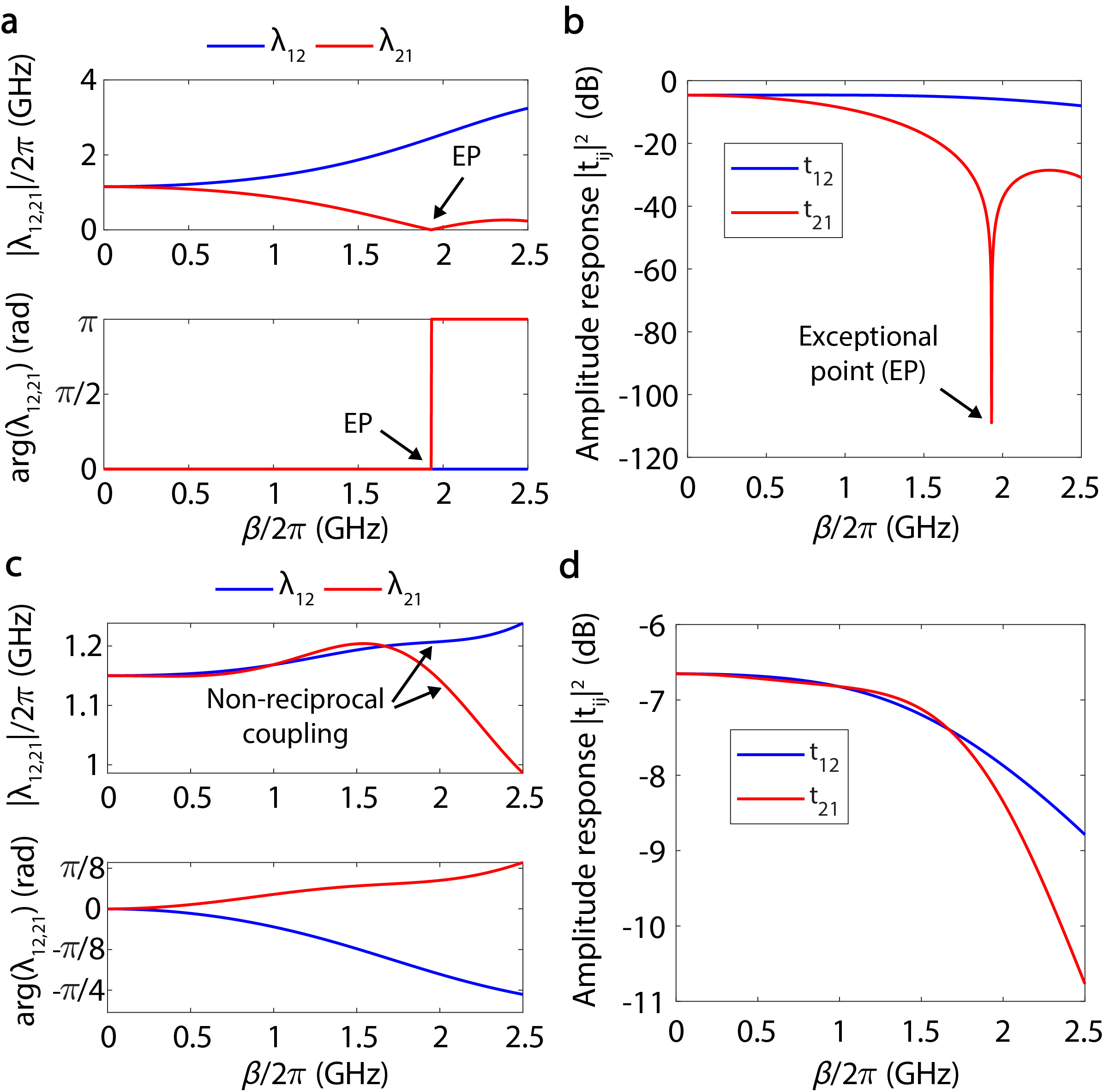}
    \centering
    \caption{
    Effective coupling strengths ($\lambda_{12,21}$) and amplitude responses ($t_{12,21}$) of both the central chain and first-order sideband at $+\omega_m$ (same at $-\omega_m$ due to symmetry) for different modulation depths ($\beta$). Central chain: \textbf{(a)} Effective coupling strengths, \textbf{(b)} Amplitude response at $\omega = \omega_0$. First-order sideband: \textbf{(c)} Effective coupling strengths, \textbf{(d)} Amplitude response at $\omega - \omega_0 = +\omega_m$.
    }
    \label{supfig:5}
\end{figure}

In the above analysis, we derived the effective Hamiltonian for the central chain by projecting the influence of pathways with non-zero potential onto $\omega_0$ to obtain an effective value for $\lambda_{12,21}$. Using this Hamiltonian, we calculated the S parameters of our photonic molecule and demonstrated that the asymmetry in the effective coupling strengths induces non-reciprocity in the system. This approach can be similarly applied to determine the effective Hamiltonians, and consequently, the S parameters for any of the sidebands at integer multiples of $\omega_m$, most easily accomplished through numerical evaluation. As a specific example, we numerically evaluate the effective coupling rates $\lambda_{12,21}$ at the sideband at $+\omega_m$ (the result is the same for the $-\omega_m$ sideband due to symmetry) as shown in Fig.~\ref{supfig:5}. We observe that asymmetric coupling also occurs at this sideband (Fig.~\ref{supfig:5}a,c). As expected, the asymmetry translates to non-reciprocity in the transmission amplitude for both the central chain (Fig.~\ref{supfig:5}b) and the first-order chains (Fig.~\ref{supfig:5}d), which can also be seen in the main manuscript Fig.~3c. For the central chain, the presence of an exceptional point (EP), where either of the effective couplings $\lambda_{12,21} \rightarrow 0$, produces a giant nonreciprocal effect, but this is absent in the first-order sideband. Finally, we note that the effective couplings $\lambda_{12,21}$ are purely real for the central chain (Fig.~\ref{supfig:5}a) but complex valued for the first-order sideband chains (Fig.~\ref{supfig:5}c). 

\begin{figure*}
    \includegraphics[width=0.8\textwidth]{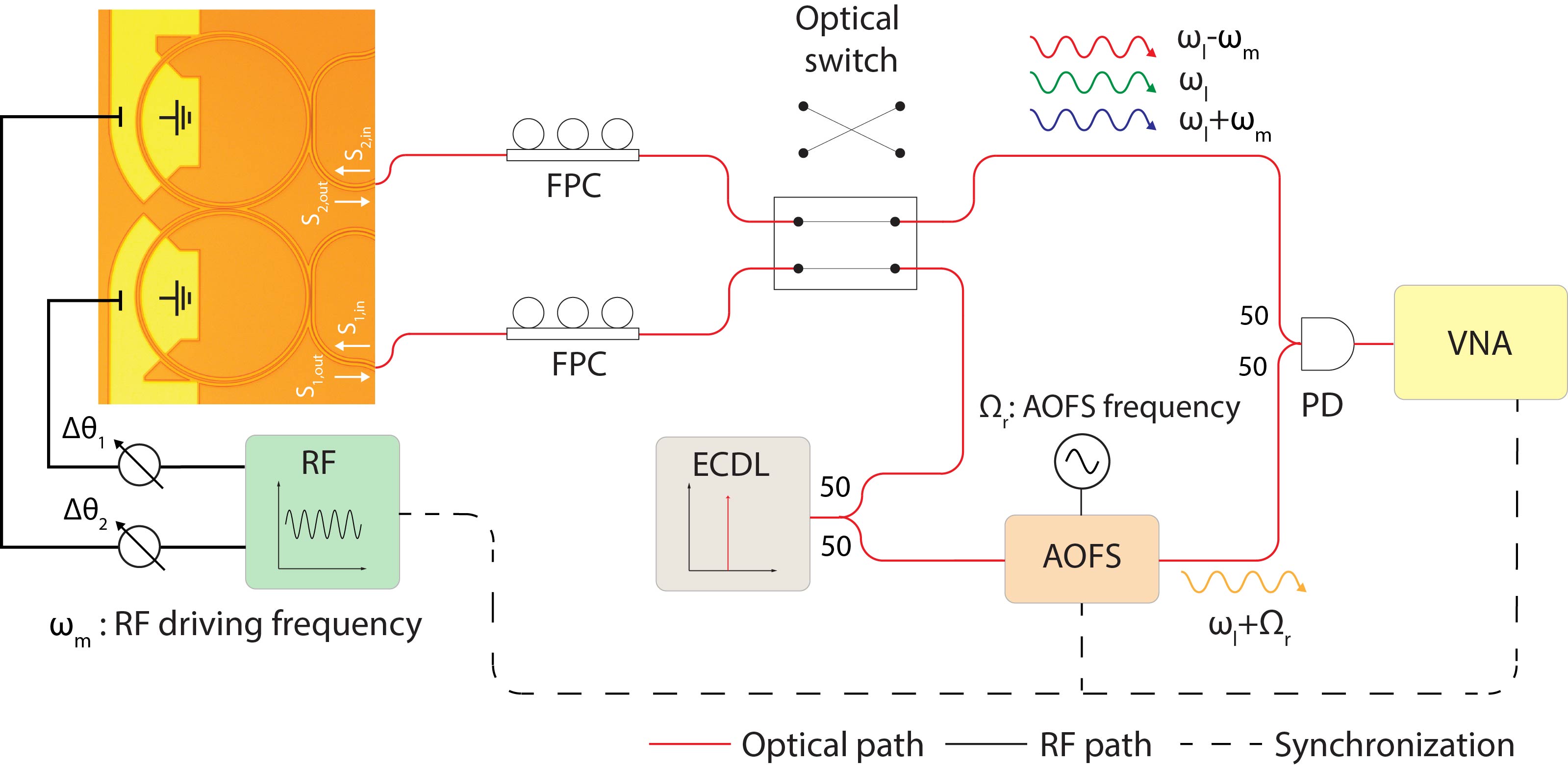}
    \centering
    \caption{
    Heterodyne detection system to measure the non-reciprocal optical transmission. ECDL: External cavity diode laser, AOFS: Acousto-optic frequency shifter, PD: Photodetector, VNA: Vector network analyzer, FPC: Fiber polarization controller
    }
    \label{supfig:2}
\end{figure*}

\section{Measurement of the Optical Transmission}
\label{sec:MeasurementTrans}

We utilize an optical heterodyne detection system, shown in Fig.~\ref{supfig:2}, to measure the carrier transmission. We use an external cavity diode laser (the New Focus model TLB-6728-P) with a narrow linewidth ($<$ 50 kHz) as the light source, which is split into two paths using a 50:50 coupler. One path is used to probe the device under test, while the other path serves as a reference. For heterodyne detection, the reference path is shifted in frequency by 100 MHz using an acousto-optic frequency shifter, and the interference of this shifted light with the probe signal is detected by a high-speed photodetector (PD). The direction of probing is controlled using an off-chip optical switch (Thorlab model OSW22-1310E). Fiber polarization controllers (FPCs) are used to adjust the polarization of the light that is coupled to the chip. The minimum per-grating coupler loss is $\approx$ 5 dB, and we factored out these losses while normalizing our experimental results. Additional information on the experimental setup and heterodyne detection method can be found in previous studies~\cite{Sohn18,Sohn_Orsel_Bahl_2021}.

\section{Measurement of the Optical Phase Response}
\label{sec:MeasurementPhase}

\begin{figure*}[th]
    \includegraphics[width=0.8\textwidth]{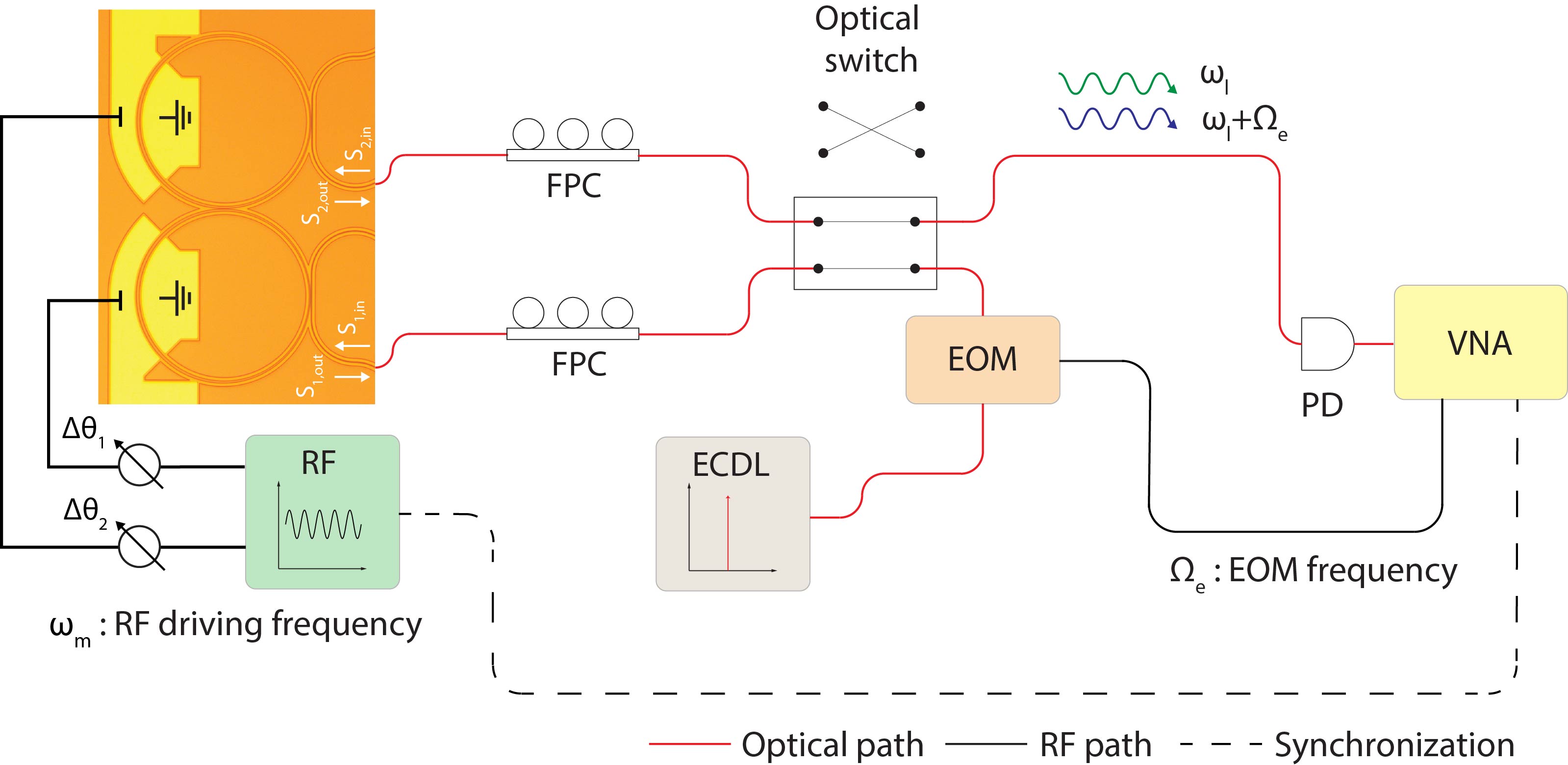}
    \centering
    \caption{
    Measurement setup with external electro-optic modulator to acquire both amplitude and phase response of the system. ECDL: External cavity diode laser, EOM: Electro-optic modulator, PD: Photodetector, VNA: Vector network analyzer, FPC: Fiber polarization controller
    }
    \label{supfig:3}
\end{figure*}

We use an external electro-optic amplitude modulator (EOM) (IML-1550-40-PMV-HER) to measure both amplitude and phase response of the optical system (Fig.~\ref{supfig:3}). Similar to the heterodyne measurement system, light (and hence the pump signal) is generated via an external cavity diode laser (the New Focus model TLB-6728-P). The probe signal is generated by applying an RF sweep to the electro-optic modulator via vector network analyzer (VNA). During the measurement, the pump signal is detuned far from the optical modes, and the probe signal scans both of the resonances due the sweep of the RF frequency. The photodetector reads the beat note signal generated by the pump and probe signals which is then fed to the VNA to measure optical transfer functions t\textsubscript{21} and t\textsubscript{12}. 

In order to calibrate our signals and remove the interference of the experimental setup, we first perform a reference optical transfer function measurement without probing any optical mode (i.e. measuring the drop transmission through the waveguide). Under this condition, the optical spectrum at the photodetector is:
\begin{multline}
s_{out}=E_Re^{-i\omega_lt}\biggl(1+\frac{\beta_{e}(\Omega_e)}{2}e^{-i(\Omega_et+\phi(\Omega_e))}\\
+\frac{\beta_{e}(\Omega_e)}{2}e^{i(\Omega_et+\phi(\Omega_e))}\biggl)+c.c
\label{eq:optical_spectrum}
\end{multline}
where $\Omega_e$ is the RF frequency applied to the external EOM, $\beta_{e}(\Omega)$ is the EOM intensity modulation coefficient while $E_R$ is the amplitude of the electric field in the device path. Here, we also defined $\phi(\Omega)$ which represents the relative delay of the sidebands with respect to the carrier. Using equation \ref{eq:optical_spectrum}, we can find the RF outputs at the photodetector as:
\begin{multline}
P_{out}=\biggl|E_Re^{-i\omega_lt}\biggl(1+\frac{\beta_{e}(\Omega_e)}{2}e^{-i(\Omega_et+\phi(\Omega_e))}\\
+\frac{\beta_{e}(\Omega_e)}{2}e^{i(\Omega_et+\phi(\Omega_e))}\biggl)+c.c\biggl|^2
\label{eq:RF_spectrum}
\end{multline}
We are interested in the signals that are at $\Omega_e$ (i.e. frequency of modulation), then the result in \ref{eq:RF_spectrum} simplifies to:
\begin{equation}
P_{\Omega_e}=4|E_R|^2\beta_{e} cos(\Omega_et+\phi(\Omega_e))
\label{eq:Real_reference_signal}
\end{equation}

While measuring the received signal, the VNA decomposes the signal in \ref{eq:Real_reference_signal} to its in-phase and quadrature components:
\begin{equation}
P_{\Omega_e}^0=4|E_R|^2\beta_{e} cos(\Omega_et+\phi(\Omega_e))cos(\Omega_et-\theta(\Omega_e))
\label{eq:In_phase_reference_signal}
\end{equation}
\begin{equation}
P_{\Omega_e}^{\pi/2}=4|E_R|^2\beta_{e} cos(\Omega_et+\phi(\Omega_e))sin(\Omega_et-\theta(\Omega_e))
\label{eq:In_phase_reference_signal}
\end{equation}
where $\theta(\Omega_e)$ is the electrical delay between the local oscillator of the VNA and the measured photodetector output. After the filtering of the components at $2\Omega_e$, the measured signals become:
\begin{equation}
P_{\Omega_e}^0=2|E_R|^2\beta_{e} cos(\phi(\Omega_e)+\theta(\Omega_e))
\label{eq:In_phase_measured_signal}
\end{equation}
\begin{equation}
P_{\Omega_e}^{\pi/2}=2|E_R|^2\beta_{e} sin(\phi(\Omega_e)+\theta(\Omega_e))
\label{eq:In_phase_measured_signal}
\end{equation}
The complex valued output signal which we use to calibrate further measurements is:
\begin{multline}
S_{1,\Omega_e}=2|E_R|^2\beta_{e}(cos(\phi(\Omega_e)+\theta(\Omega_e)\bigl)\\
+isin\bigl(\phi(\Omega_e)+\theta(\Omega_e)))=4|E_R|^2\beta_{e} e^{i(\phi(\Omega_e)+\theta(\Omega_e))}
\label{eq:Complex_reference_signal}
\end{multline}
We now can acquire the complex response of the photonic molecule by measuring the through transmission. In this case one of the generated sidebands vanishes (it is the Stokes sideband for this case) due to the lack of the optical states, and the RF outputs at the photodetector becomes:
\begin{equation}
P_{out}=\left|E_Re^{-i\omega_lt}\left(t_c+t_{us}\frac{\beta_{e}(\Omega_e)}{2}e^{-i(\Omega_et+\phi(\Omega_e))}\right)+c.c\right|^2
\label{eq:RF_spectrum_through}
\end{equation}
Here, we used the variables $t_c$ and $t_{us}$ to represent the complex transmission coefficients of the carrier and the upper sideband respectively. Ideally, $t_c=t_c^*$ with $|t_c|<1$ so that we can retrieve the true cavity response. Then, the measured signal at $\Omega_e$ becomes:
\begin{equation}
P_{\Omega_e}=2|E_R|^2\beta_{e}|t_c||t_{us}|cos(\Omega_et+\phi(\Omega_e)+\Theta_c(\Omega_e)+\Theta_{us}(\Omega_e))
\label{eq:Real_signal}
\end{equation}
Here, $\Theta_c$ and $\Theta_{us}$ are the phase response of the carier and the sideband respectively. After the measurement using VNA, the complex signal becomes:
\begin{multline}
S_{2,\Omega_e}=|E_R|^2\beta_{e}|t_c||t_{us}|(cos(\phi+\theta+\Theta_c+\Theta_{us})\\
+isin(\phi+\theta+\Theta_c+\Theta_{us}))=2|E_R|^2\beta_{e}|t_c||t_{us}| e^{i(\phi+\theta+\Theta_c+\Theta_{us})}
\label{eq:Complex_signal}
\end{multline}
We can find the response of the sideband signal and hence the cavity response by dividing eq.~\ref{eq:Complex_signal} with eq.~\ref{eq:Complex_reference_signal}:
\begin{equation}
\frac{S_{2,\Omega_e}}{S_{1,\Omega_e}}=\frac{|t_c||t_{us}|e^{i(\Theta_c+\Theta_{us})}}{2}
\label{eq:Calibrated_signal}
\end{equation}
Then:
\begin{equation}
t_{us}=\left(\frac{2S_{2,\Omega_e}}{t_cS_{1,\Omega_e}}\right)
\label{eq:True_response}
\end{equation}
The expression $t_{us}$ ($t_{pq}$ in the main manuscript) now represents the calibrated cavity response. During our experiments, we positioned our carrier signal so that $|t_c|$ is purely real (i.e. the carrier signal does not probe the photonic molecule), and we used eq.~\ref{eq:True_response} to measure both the phase and amplitude response of the system for both directions.

\end{document}